\documentclass[preprint,showpacs,preprintnumbers,amsmath,amssymb]{revtex4}

\usepackage{color,vmargin,epsfig,rotating,graphicx,subfigure}
\usepackage{amsmath,amssymb,amscd}
\usepackage[latin1]{inputenc}
\usepackage{dsfont}

\setpapersize{A4}

\renewcommand{\Re}{{\rm Re}}
\renewcommand{\Im}{{\rm Im}}

\newcommand{\ri}{{\rm i}}
\newcommand{\re}{{\rm e}}
\newcommand{\rd}{{\rm d}}

\begin{document}

\title{Thermal heat radiation, near-field energy density and near-field radiative heat transfer 
       of coated materials}

\author{Svend-Age Biehs}

\affiliation{Institut f\"ur Physik, Carl von Ossietzky Universit\"at,
        D-26111 Oldenburg, Germany}

\date{May 29, 2007}

\pacs{44.40.+a, 78.66.-w, 05.40.-a, 41.20.Jb}

\begin{abstract}
We investigate the thermal radiation and thermal near-field energy 
density of a metal-coated semi-infinite body for different substrates. We show that the surface polariton coupling within
the metal coating leads to an enhancement of the TM-mode part of the thermal near-field energy density when a polar 
substrate is used. In this case the result obtained for a free standing metal film is retrieved. In contrast, in the case of
a metal substrate there is no enhancement in the TM-mode part, as can also be explained within the framework
of surface plasmon coupling within the coating. Finally, we discuss the influence of the enhanced thermal energy density 
on the near-field radiative heat transfer between a simple semi-infinite and a coated semi-infinite body for different 
material combinations.
\end{abstract}

\maketitle

\section{Introduction}
\label{Sec1}

The fluctuating electrodynamic near field close to the surface of dielectric bodies due to 
thermal and quantum fluctuations inside that bodies has come to the fore 
in the last decade. The growing interest of researchers
in the investigation of fluctuating near fields is accompanied by manifold
new possibilities to measure the interesting properties of such thermal near 
fields~\cite{ShchegrovEtAl00,CarminatiGreffet99,C.HenkelEtAl2000,J.J.GreffetEtAl2002,J.J.GreffetEtAl2003,F.MarquierEtAl2004,K.JoulainEtAl2005}, 
which have been developed
in the last decade. For example, two of these techniques are the thermal radiation scanning tunneling
microscopy~\cite{Y.DeWileEtAl2006} and the usage of Bose-Einstein condensates~\cite{J.M.ObrechtEtAl2007}. 
Moreover, near field scanning thermal microscopy~\cite{KittelEtAl05} (NSThM) is a new possibility to measure the
radiative heat transfer, which is itself related to the 
properties of the fluctuating near field~\cite{Dorofeyev98,MuletEtAl01,KittelEtAl05} between dielectric 
bodies~\cite{PolderVanHove71}. From the experimental point of view it should also
be helpful to study the near field of coated materials.

Not only the significance of the fluctuating near field in scanning probe techniques or 
nanotechnological applications makes a theoretical investigation necessary and useful. 
The electrodynamic near field is also of great theoretical interest, because it shows new 
and unexpected physical properties. For example, it has been shown in 
recent publications~\cite{CarminatiGreffet99} that coherent quasi-monochromatic evanescent waves can exist in 
the thermal near field, although the latter is generated by fluctuating thermal sources.
In order to study near-field effects one may calculate different physical quantities such
as the cross-correlation tensor~\cite{CarminatiGreffet99}, the local density of 
states (LDOS)~\cite{JoulainEtAl03} or the spectral energy density~\cite{ShchegrovEtAl00} 
in the vicinity of the dielectric body, where this body is usually assumed to be a semi-infinite medium.

\begin{figure}[Hhbt]
  \centering
  \begin{minipage}[t]{0.9\textwidth}
    \epsfig{file=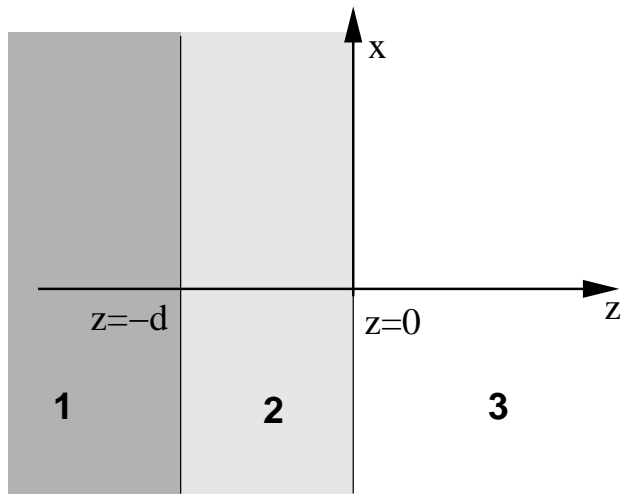, width = 0.6\textwidth}
    \caption{Sketch of the configuration used here: The bulk extends in the regime $z \leq -d$, the coating ranges from
             $z = -d$ to $z=0$, and the half-space $z > 0$ is assumed to be vacuum.}
    \label{Fig:Configuration}
  \end{minipage}
\end{figure}

In this paper we will study how a coating influences the thermal electrodynamical
near field of a semi-infinite substrate (see fig.\ \ref{Fig:Configuration}). For that reason we calculate the energy density
above the coated body. We will show that an effect predicted for a free standing metallic film~\cite{Biehs2007} 
can be retrieved by using a polar material as substrate, whereas for a metal substrate the thermal
near-field energy density changes dramatically. Both cases are discussed and understood with the help of surface
plasmon coupling within the coating. Furthermore we investigate how the coating on different substrates
influences the near-field radiative heat transfer. 

The near-field radiative heat transfer was already disussed in such a slab configuration for
polar materials~\cite{A.NarayanaswamyC.Chen2003} for an application in thermophotovoltaics and briefly for 
metal substrates coated with different metals~\cite{A.I.VolokitinB.N.J.Persson2001}. Here we explicitly
calculate the near-field radiative heat transfer between a semi-infinite and a coated semi-infinite body, showing
in detail that the physical mechanisms leading to different energy densities will leave their imprints in the near-field
radiative heat transfer. With this information at hand it could for example be possible to give a better understanding
of the signal measured with a NSThM, and to clarify the question whether that signal can be interpreted within a 
dipole-model~\cite{Dorofeyev98,MuletEtAl01} 
or whether it can be modelled as the heat transfer between a semi-infinite and a coated semi-infinite body. Furthermore, the results derived in 
this and the preceding paper~\cite{Biehs2007} can also serve as a basis for the investigation 
of near-field effects for coated materials, which are often used in experimental setups.

This paper is a direct follow-up of reference~\cite{Biehs2007}, so we refer the reader to that paper for a brief discussion 
of Rytov's fluctuational electrodynamics~\cite{RytovEtAl89}. When considering the geometry given in 
fig.\ \ref{Fig:Configuration}, it is in principle necessary to construct the dyadic Green's 
function with observation point located in the regime $z > 0$, and with sources within the coating or the substrate, respectively.
Since we already determined the dielectric Green's 
function for a dielectric film, as corresponding to the coating, in all details~\cite{Biehs2007}, the dyadic Green's function with  
source currents within the coating can directly be taken from that reference. 
The dyadic Green's function with
sources within the substrate can be constructed
in a straightforward way, so we present here only the results and 
refer the interested reader to~\cite{Biehs2007} and~\cite{ChenToTai71}, respectively. For convenience we use
the same notation as in our preceding paper~\cite{Biehs2007}, and for comparability we use again for numerical
computations the Drude model for metals and the Reststrahlen formula for polar materials with material
parameters taken from~\cite{AshcroftMermin76,S.Adachi2004}.

This paper is organized in the following way: In section \ref{Sec2} we briefly discuss the thermal radiation of a coated material.
In section \ref{Sec3} we study the thermal near field of the coated material for different coatings and substrates and show in section
\ref{Sec4} how the observed effects can be interpreted with the surface plasmon polariton coupling inside the coating. Finally,
in the last section we calculate the near-field radiative heat transfer and discuss the influence of a metal coating.

\section{Thermal radiation}
\label{Sec2}

In this paper we are mainly interested in the evanescent near field
of the coated semi-infinite body, but for the sake of completeness we
also report the results for the radiative part. In order to derive
the thermal radiation of the coated semi-infinite body we calculate
the averaged $z$-component $\langle S_z \rangle$ of the Poynting vector outside the 
layered system in fig.\ \ref{Fig:Configuration}, which is assumed to be
in local thermal equilibrium at temperature $T$, setting $\epsilon_3 = \epsilon_0$. Taking fluctuating
source currents inside the bulk medium (the substrate) with permittivity $\epsilon_1$ 
and inside the coating with permittivity $\epsilon_2$, which 
contribute additively to the Poynting vector outside the layered system, 
we get after a lengthy but straightforward calculation
\begin{equation}
  \langle S_z \rangle = \int\!\!{\rm d} \omega\, \frac{E(\omega, \beta)}{(2 \pi)^2} \int\!\!\!{\rm d} \lambda \, 
                            \lambda {\rm e}^{- 2 h_0'' z} \bigl( T_\perp^{{\rm total}} + T_\parallel^{{\rm total}}\bigr).
\label{Eq:Poynting_vector}
\end{equation} 
The transmission coefficents $T^{{\rm total}}$ are given as the sum of the bulk and coating
transmission coefficients, $T^{{\rm b}} + T^{{\rm c}}$, for TM- and TE-polarization ($\parallel$ and $\perp$), respectively. 
The transmission coefficients for the bulk contribution are given by
\begin{align}
  T^{{\rm b}}_\perp     &= 16 |h_2|^2 \frac{\Re(h_0) \Re(h_1)}{|D_\perp|^2}, \nonumber \\
  T^{{\rm b}}_\parallel &= 16 |h_2|^2 \frac{|k_2|^4}{|k_1|^4} \frac{\Re(h_0) \Re(h_1 \overline{\epsilon}_{1})}{|D_\parallel|^2} 
  \label{Eq:Transmission_Bulk}
\end{align}
with $h_i = \sqrt{k_0^2 \epsilon_i - \lambda^2}$ for $i = 0, 1, 2$ and
\begin{equation}
  D = a^{12} a^{02} {\rm e}^{-{\rm i} h_2 d} - b^{12} b^{02} {\rm e}^{{\rm i} h_2 d}.
\end{equation} 
The coefficients $a$ and $b$ are defined as
\begin{align}
  a^{ij}_\perp &:= h_i + h_j \label{a_te},\\
  a^{ij}_\parallel &:= h_i \frac{\epsilon_j}{\epsilon_i} + h_j \label{a_tm},\\
  b^{ij}_\perp &:= h_i - h_j \label{b_te},\\
  b^{ij}_\parallel &:= h_i \frac{\epsilon_j}{\epsilon_i} - h_j.
  \label{b_tm}
\end{align}
The transmission coefficients for the coating have already been calculated in~\cite{Biehs2007}
and can be stated as
\begin{align}
  T_\perp^{{\rm c}}     &= \frac{4 \Re(h_0)}{|D_\perp|^2} \biggl[ \Re(h_2) A_\perp
                            + 2 \Im(h_2) B_\perp\biggr]  \nonumber \\
  T_\parallel^{{\rm c}} &= \frac{4 \Re(h_0)}{|D_\parallel|^2} \biggl[ \Re(h_2 \overline{\epsilon}_{r2}) A_\parallel
                            + 2 \Im(h_2 \overline{\epsilon}_{r2}) B_\parallel\biggr]
  \label{Eq:Transmission_Coating}
\end{align}
with
\begin{align}
  A &= |a^{12}|^2 \bigl({\rm e}^{2 h_2'' d} - 1 \bigr) 
       \quad- |b^{12}|^2 \bigl({\rm e}^{- 2 h_2'' d} - 1\bigr),  \\
  B &=  \Im\biggl(a^{12} \overline{b^{12}} \bigl({\rm e}^{- 2 {\rm i} h_2' d} - 1\bigr)\biggr), 
  \label{Eq:Definition_of_Coefficients}
\end{align}
where we have used the notation $h_i = h_i' + \ri h_i''$.
Even though the transmission coefficients, which are rather complicated, 
could be reformulated in term of Fresnel reflection coefficients~\cite{J.D.Jackson1999},
we will not perfom this procedure here, because in that case we 
get different forms of transmission coefficients for the propagating and evanescent modes (cf.\ \cite{Biehs2007}),
i.e., we get four equations instead of the two given in 
(\ref{Eq:Transmission_Bulk}) and (\ref{Eq:Transmission_Coating}), thus unnecessarily
inflating the formalism. But it should be kept in mind that the transmission coefficients,
which can be stated with one equation for the propagating part with $\lambda < k_0$
and the evanescent part with $\lambda > k_0$, behave in a quite different manner
for propagating and evanescent modes, respectively. This is a consequence of
the fact that $h_0$ is purely real for propagating modes or purely imaginary for evanescent modes,
$h_0=\ri\sqrt{\lambda^2 - k_0^2}\equiv\ri\gamma$. Therefore, the evanescent component $T^{{\rm total}}_{\rm ev}$
does not contribute to the expression for the Poynting vector (\ref{Eq:Poynting_vector}), i.e.,
the Poynting vector covers information on the propagating modes only. 

\begin{figure}[Hhbt]
  \centering
  \begin{minipage}[t]{0.9\textwidth}
    \epsfig{file=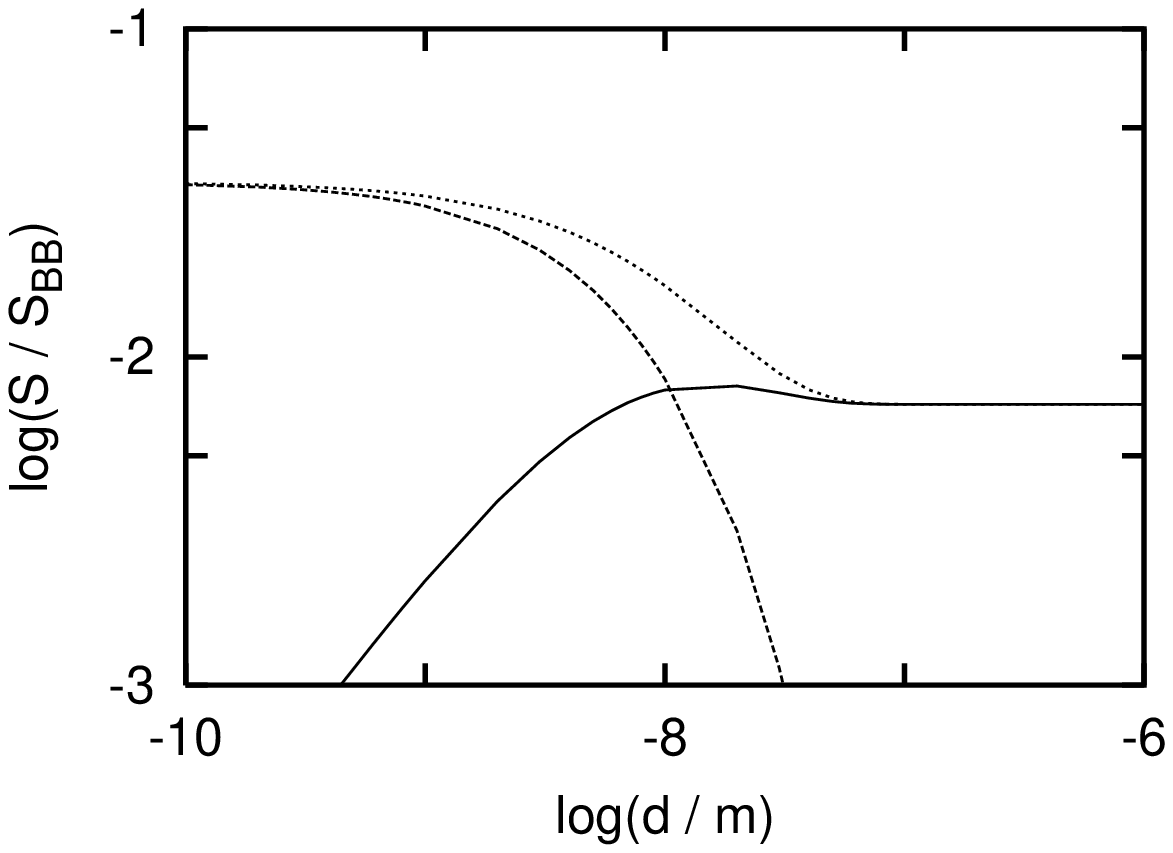, width = 0.45 \textwidth}
    \epsfig{file=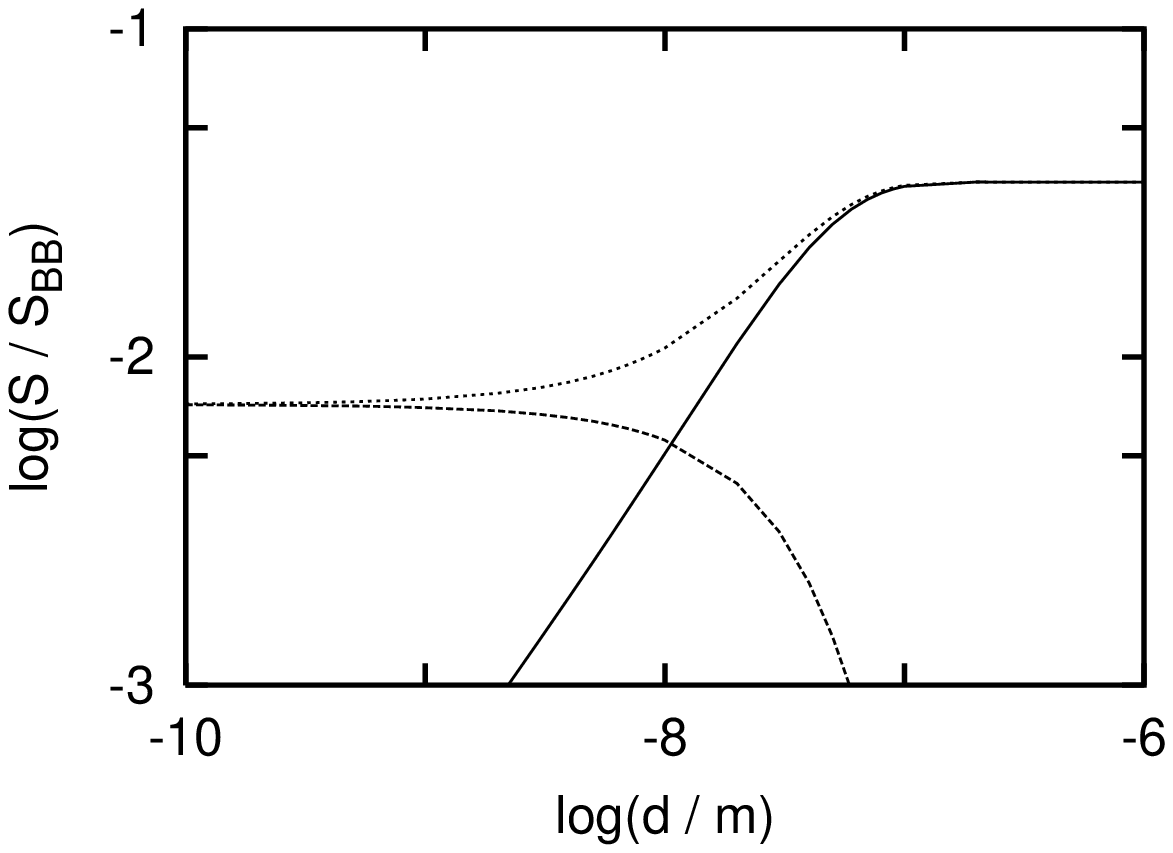, width = 0.45 \textwidth}
    \caption{Left: Numerical result for the thermal radiation of a Pt-coated Au-substrate at temperature $T = 300 {\rm K}$
             for different thickness $d$ of the coating, normalized to the black body value $S_{\rm BB}$ given
             by the Stefan-Boltzmann law. The solid line is the contribution of the coating and
             the dashed line that of the substrate, whith the sum of both being given by the dotted line. 
             Right: Here the role of the substrate and coating
             are interchanged, so that this panel shows the thermal radiation of a Au-coated Pt-substrate at temperature $T = 300 {\rm K}$.}
    \label{Fig:Poynting_Pt_Au}
  \end{minipage}
\end{figure}
 
Before we present numerical results for the Poynting vector, 
we specify the limiting values of the transmission coefficients
for the propagating modes for different layer thickness $d$, considering
the two cases $d \gg 1/h_2''$ and $d \ll 1/h_2''$, i.e., layers much thicker
or thinner than the skin depth of the coating material, given by 
\begin{equation}
  d_{\rm s} = \frac{1}{k_0 \Im(\sqrt{\epsilon_r2})} \approx \frac{1}{h_2''}.
\label{Eq:Skin_depth}
\end{equation}
For thin coatings with $d \ll d_s$ the
transmission coefficients $T^{\rm c}$ go linearly with thickness $d$ 
to zero (cf.\ \cite{Biehs2007}), whereas the transmission coefficents of the bulk 
$T^{\rm b}$ converge to the transmission coeffient of a semi-infinite 
body~\cite{PolderVanHove71} with permittivity $\epsilon_1$, i.e.,
\begin{align}
  T^{{\rm total}}_\perp &\rightarrow T^{{\rm b}}_\perp \approx 4 \frac{\Re(h_1) \Re(h_0)}{|a_\perp^{01}|^2}, \nonumber \\
  T^{{\rm total}}_\parallel &\rightarrow T^{{\rm b}}_\parallel \approx  4 \frac{\Re(h_1 \overline{\epsilon}_{r1}) \Re(h_0)}{|a_\parallel^{01}|^2}.
  \label{Eq:Semi-infinite}
\end{align}
In contrast, for thick coatings with $d \gg d_s$ the transmission coefficents 
of the bulk contributions go to zero and the transmission coefficients of the
coating converge to the transmission coeffient of a semi-infinite 
body~\cite{PolderVanHove71} with permittivity $\epsilon_2$, which can 
be derived from eq.\ (\ref{Eq:Semi-infinite}) by exchanging the index 1 with 2.
  
Therefore, the thermal radiation of a coated body given by propagating
modes only and being independent of $z$ (because $h_2'' = 0$ for propagating modes),
has different values for different thicknesses $d$ of the coating. In the limit that the coating
is very thick, i.e., $d \gg d_{\rm s}$, the radiation is that of a half-space filled with the coating
material only. In the other limit of very thin coating, i.e.\ $d \ll d_{\rm s}$, the radiation is that of
a half-space filled entirely with the bulk material. In general, the value of the Poynting vector always falls
between these two extremes. Thus it seems that the thermal radiation maximum found for free standing metallic films 
of a certain thickness~\cite{Biehs2007} cannot be observed for coated materials.
This is illustrated in the left panel of fig.\ (\ref{Fig:Poynting_Pt_Au}), where there is a maximum in the contribution
of the coating, but this is overlayed by the bulk contribution.

\section{Thermal near field}
\label{Sec3}

Next, we discuss the non-radiative part of the fluctuating near field
in the vicinity of the coated substrate. To this end, we investigate
the energy density in the distance $z$ from the coated body, which can be written as
\begin{equation}
  \langle u(z) \rangle = \int\!\!\rd\omega\,\frac{E(\omega,\beta)}{(2 \pi)^2} \int\!\!\rd\lambda\, \lambda \frac{\lambda_s^2}{2 \omega} \re^{- 2 h_0'' z} \frac{\bigl(T_\perp^{{\rm total}} + T_\parallel^{{\rm total}} \bigr)}{\Re(h_0)},
  \label{Eq:Energy_density}
\end{equation}
with $\lambda_s^2 = 2 k_0^2$ for propagating modes with $\lambda < k_0$ and 
$\lambda_s^2 = 2 \lambda^2$ for evanescent modes with $\lambda > k_0$.
Here the factor $\Re(h_0)$ appearing in the transmission coefficients (\ref{Eq:Transmission_Bulk}) 
and (\ref{Eq:Transmission_Coating}) is canceled out by the 
denominator in eq.\ (\ref{Eq:Energy_density}), so that the energy
density contains information about the evanescent thermal near field. Due to these 
evanescent modes the expression for the energy density becomes dependent on the distance $z$ from the layered system, 
although the contribution of the propagating modes is again independent of the observation distance $z$. 

\begin{figure}[Hhbt]
  \centering
  \begin{minipage}[t]{0.9\textwidth}
    \epsfig{file=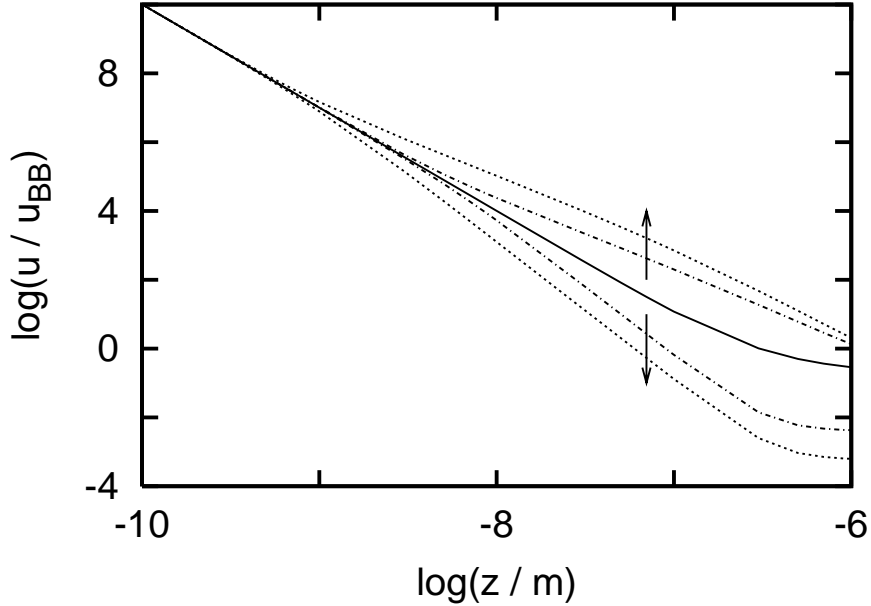, width = 0.9 \textwidth}
    \caption{Numerical results for the thermal near-field energy density  $\langle u^{\rm c}_\parallel \rangle$ of a Bi-coating
             on a GaN- and an Al-substrate with temperature $T = 300 {\rm K}$, as functions of 
             the observation distance $z$ from
             the layered system, normalized to the corresponding black body value. 
             We plot here the results for different thicknessess $d$ of the coating material, with 
             the solid line giving the thermal energy density above a semi-infinite Bi medium. The dashed lines
             give the TM-mode part of thermal energy density for $d = 5\cdot10^{-9} {\rm m}$, 
             and the dotted lines for $d = 1\cdot10^{-9} {\rm m}$. As indicated by the arrows,
             for the case of the polar substrate GaN the energy density raises over that of the 
             semi-infinite Bi medium. On the other hand, the energy density for the metal substrate Al is diminished in
             comparison to that of the semi-infinite Bi medium at distances $z \gg d$.  }
    \label{Fig:Dens_Bi_auf_GaN_bzw_Al_tm}
  \end{minipage}
\end{figure}

Taking the limits for thin and thick coatings is in this case not easy, because for 
the evanescent modes the transmission coefficients
of the coating $T^{\rm c}$ contain expressions depending 
on $h_2'' d$ in the nominator and denominator which compete with each other, in
analogy to the behaviour discussed in ref.~\cite{Biehs2007} for a single thin dielectric layer.
In contrast, the limit of the transmission coefficients $T^{\rm b}$ for a thin coating with $h_2'' d \ll 1$
reduces for both propagating and evanescent modes to the expression (\ref{Eq:Semi-infinite}) and vanishes
for thick coatings, $h_2'' d \gg 1$.  

\begin{figure}[Hhbt]
  \centering
  \begin{minipage}[t]{0.9\textwidth}
    \epsfig{file=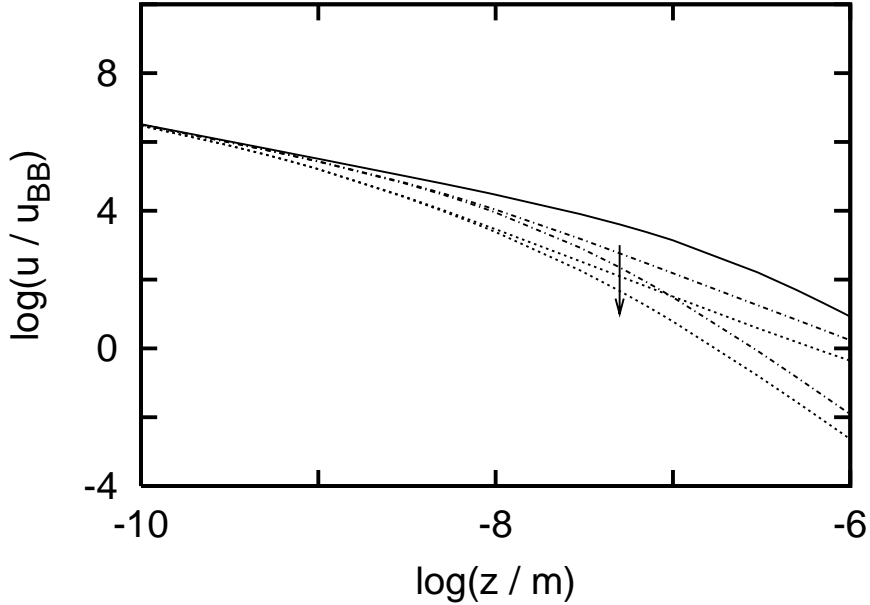, width = 0.9 \textwidth}
    \caption{As fig.\ \ref{Fig:Dens_Bi_auf_GaN_bzw_Al_tm} but for $\langle u^{\rm c}_\perp \rangle$, i.e., for
             the TE-mode contribution. Symbols are as in fig.\ \ref{Fig:Dens_Bi_auf_GaN_bzw_Al_tm}.
             In this case the thermal energy density of the coatings falls below that of a semi-infinite 
             Bi medium for $d \ll d_{\rm s}$, such that the thermal energy density $\langle u^{\rm c}_\perp \rangle$
             obtained with a metal substrate is smaller than that for a polar substrate at distances $z \gg d$.}
    \label{Fig:Dens_Bi_auf_GaN_bzw_Al_te}
  \end{minipage}
\end{figure}

In the evanescent-mode regime $\lambda > k_0$ the 
energy density depends on $z$ or, more precisely, on $\exp(- 2 h_2'' z)$. 
From this fact and the form of the transmission coefficients given in eq.\ (\ref{Eq:Transmission_Coating}) 
it appears reasonable to discuss the cases of thin and thick coatings, i.e., $h_2'' d \gg 1$ 
and $h_2'' d \ll 1$, in the regions $z \ll d$ and $z \gg d$ separately. 
As follows from ref.~\cite{Biehs2007}, in the region $z \ll d$ the transmission
coefficients $T^{\rm c}$ take the same form as those for a half-space filled entirely with the coating material.
It can be shown that for $z \ll d$ the bulk contribution $T^{\rm b}$ becomes negligible. 
This is a reasonable result, because the evanescent waves with the lateral wave vector $\lambda$
are damped at a length scale $\lambda z \approx 1$ above the layered system. Therefore for $z \ll d$
the near field is dominated by evanescent waves with $\lambda^{-1} \approx z \ll d$,
which do not carry information about the restriction due to the 
finite layer thickness $d$. From that it seems to be clear that for $z \ll d$
one receives a result which coincides with that for a bulk made up of the
coating material, i.e., with the permittivity $\epsilon_2$. 

\begin{figure}[Hhbt]
  \centering
  \begin{minipage}[t]{0.4\textwidth}
    \epsfig{file=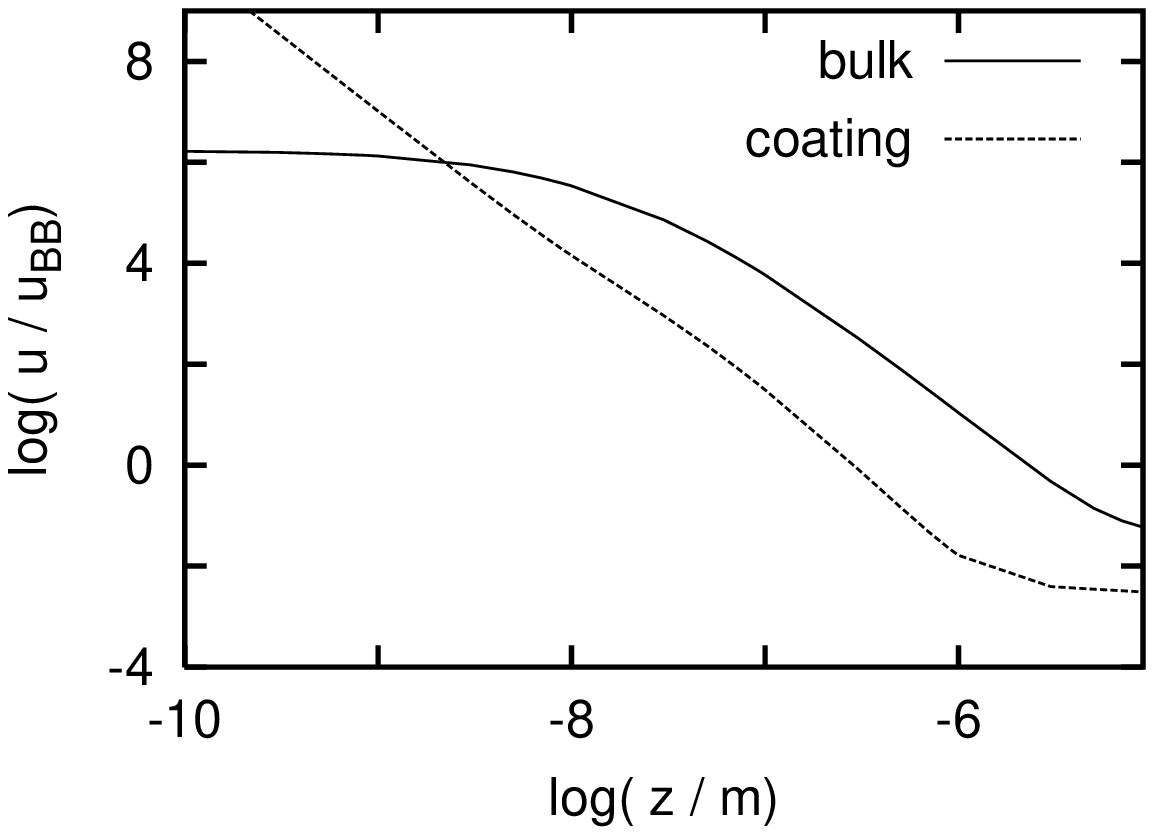, width = 0.9 \textwidth}
  \end{minipage}
  \begin{minipage}[t]{0.4\textwidth}
    \epsfig{file=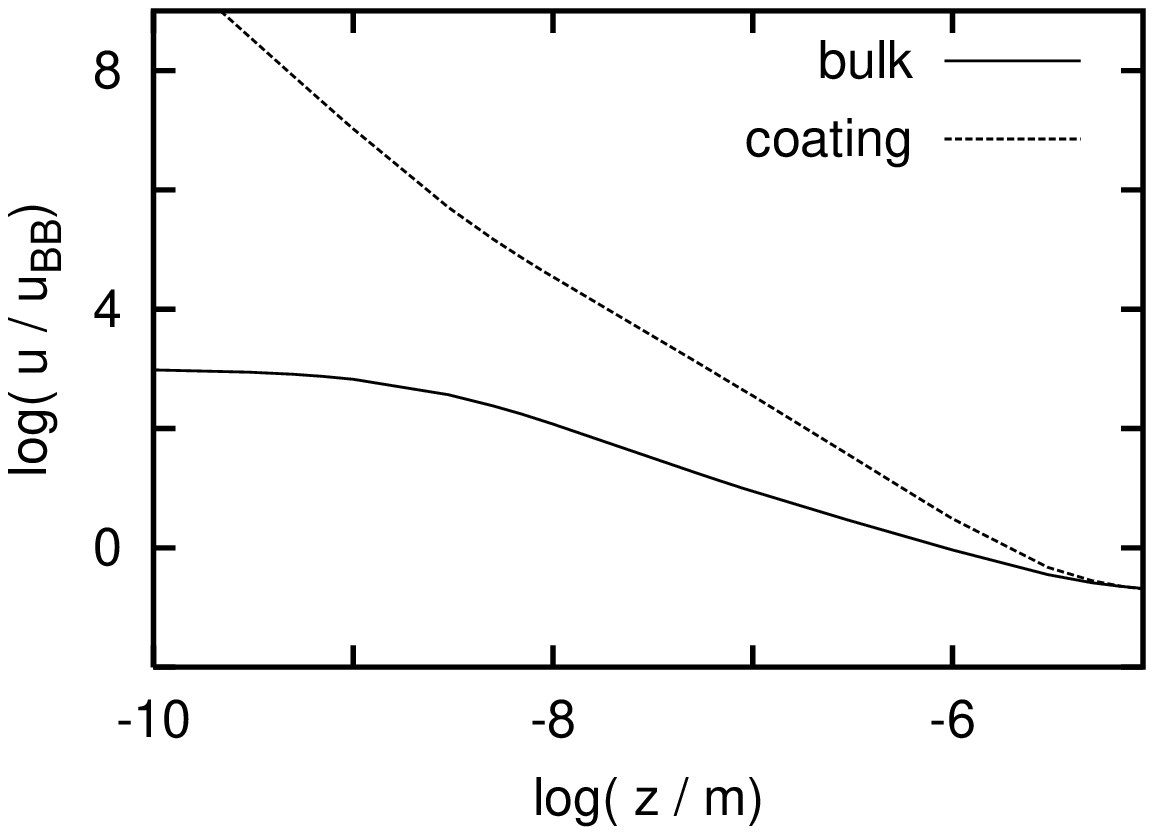, width = 0.9 \textwidth}
  \end{minipage}
  \caption{Numerical results for the thermal near-field energy density $\langle u^{\rm c} \rangle$  and $\langle u^{\rm b} \rangle$
           for a $5 {\rm nm}$ Bi-coating on a Al- (left) and GaN- (right) substrate, assuming $T = 300 {\rm K}$.
           It is seen that for a thin coating with $d \ll d_{\rm s}$ on a metal substrate the energy density 
           above the layered structure
           can be dominated by the contribution of the substrate for distances $z \gg d$, whereas this conclusion cannot be drawn
           for a polar substrate.}
  \label{Fig:Dens_Bi_auf_GaN_bulk_coating}
\end{figure} 

For $d \ll d_{\rm s}$ and $z \gg d$ the situation is more complex, as far as TM modes are concerned. 
The TM-mode contribution to the thermal energy density given by the coating is in that region 
given by~\cite{Biehs2007}
\begin{equation}
  \langle u^{{\rm c,ev}}_\parallel \rangle \approx \int\!\!\rd\omega\, \frac{E(\omega,\beta)}{(2 \pi)^2} \frac{2}{z^3 \omega} 
           \int\!\!\rd \eta\, \eta^2 \frac{\Im(r^{02}_\parallel) \re^{-2 \eta}}{|1 - r^{12}_\parallel r^{02}_\parallel (1 - 2 \eta \frac{d}{z})|^2} 
           \bigl[ 2 \eta \frac{d}{z} (1 + |r^{12}_\parallel|^2) \bigr],
  \label{Eq:Energy_density_coating}
\end{equation}
where $r^{12}$ and $r^{02}$ are the usual Fresnel reflection coefficients~\cite{J.D.Jackson1999} for the interfaces
at $z = - d$ and $z = 0$, respectively, and $\eta \equiv \lambda z$.
Through these reflection coefficients, the energy density depends on the properties of bulk and coating material.
Let us restrict the following discussion to metal coatings, so that $|r^{02}| \approx 1$. 
Now the energy density contribution of the 
coating solely depends on the choice of bulk material. If we choose
as bulk material the vacuum or a polar material, i.e., $r^{12} = r^{02}$ or $r^{12} \approx r^{02}$, 
then the expression for the energy density reduces to
\begin{equation}
  \langle u^{{\rm c,ev}}_\parallel \rangle \approx \int\!\!\rd\omega\, \frac{E(\omega,\beta)}{(2 \pi)^2} 
                                 \frac{4}{z^2 d \omega} \int\!\!\rd \eta\, \eta \, \Im(r^{02}_\parallel) \re^{-2 \eta}.
\end{equation}
For a metal film or a coated polar material, respectively, we get a 
$1/z^2$-dependence of the energy density, as discussed in~\cite{Biehs2007}.
(For a metal coating obeying the Hagens-Rubens approximation the power laws
derived in~\cite{Biehs2007} also give reasonable approximations for a polar substrate.)
In contrast, if we take a second metal as bulk material, then the 
reflection coefficients $r^{12}$ between these two metals should be small, so
we can approximate the denominator in eq.\ (\ref{Eq:Energy_density_coating})
by 1. As a consequence we find a $1/z^4$-dependence of the energy
density for the TM-modes for coated metals. Therefore, the $1/z^3$-dependence
of $\langle u_\parallel \rangle$ provided by a half-space consisting solely of 
the coating material for $z \ll d$ changes to a $1/z^2$- or $1/z^4$-dependence
for $z \gg d$ when considering a polar or metal bulk with a metal coating. In fig.\ 
\ref{Fig:Dens_Bi_auf_GaN_bzw_Al_tm} this splitting is shown for a Bi-coating 
of different thicknesses $d$ on a GaN bulk and an Al bulk, respectively.
It is interesting to see that the contribution of the coating material 
to the energy density $\langle u_\parallel^{{\rm c}} \rangle$ for polar bulk
materials becomes greater than its bulk value for distances $z \gg d$,
similar to what has been discussed for thin metall films in ref.~\cite{Biehs2007}. 

\begin{figure}[Hhbt]
  \centering
  \begin{minipage}[t]{0.9\textwidth}
    \epsfig{file=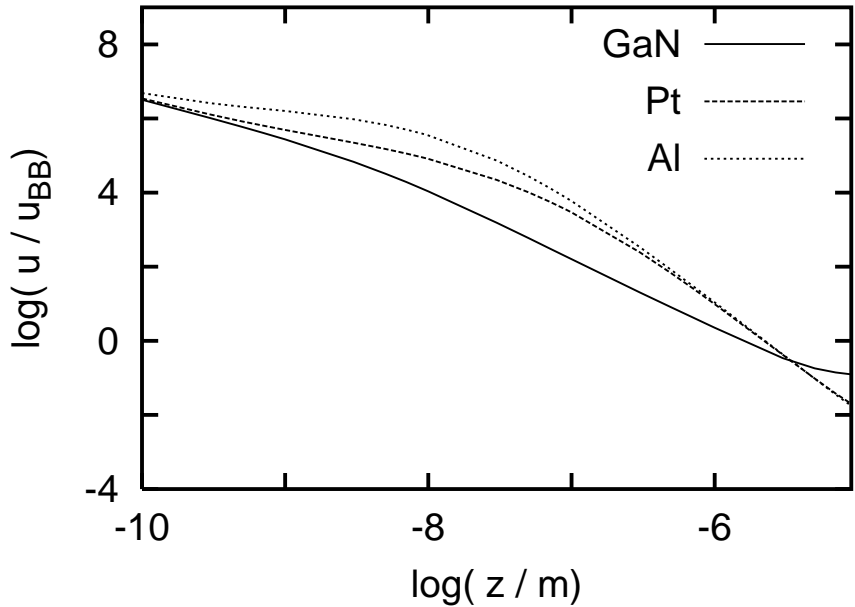, width = 0.9 \textwidth}
    \caption{Numerical results for the total thermal energy density $\langle u^{\rm total}_\perp \rangle$ 
             above the layered structure for a substrate consisting of GaN, Pt or Al coated with a $5 {\rm nm}$ layer
             of Bi at a temperature $T = 300 \,{\rm K}$.}
    \label{Fig:Dens_Bi_te}
  \end{minipage}
\end{figure}

Such a splitting can also be observed for the TE-mode part of the energy density
contribution of the coating material $\langle u_\perp^{{\rm c}} \rangle$ for 
distances $z \gg d$, as shown in fig.\ \ref{Fig:Dens_Bi_auf_GaN_bzw_Al_te}. 
But in contrast to $\langle u_\parallel^{{\rm c}} \rangle$ the energy density
of the coating material does never rise over its bulk value. From the numerical
result displayed in fig.\ \ref{Fig:Dens_Bi_auf_GaN_bzw_Al_te} one can infer that for a
coated polar bulk material $\langle u_\perp^{{\rm c}} \rangle$ again has
a $1/z^2$-dependence for $z \gg d$, whereas for coated metals there seems to be no
well-developed power law. 

Now let us study the interplay of the contributions of the bulk or substrate and
that of the coating to the thermal energy density for $d \ll d_{\rm s}$. 
From the discussion above it follows
that for $z \ll d$ it is always $\langle u^{\rm c} \rangle$ which dominates the total 
energy density, with the value of $\langle u^{\rm c} \rangle$ coinciding
with its half-space value, i.e., being independent of the coating thickness $d$.  
For distances $z \gg d$ it is {\itshape a priori} not clear whether the bulk or the coating
contribution dominates the energy density. However, one may expect for a
polar bulk material and a metal coating that the bulk contribution does not
play an important role because $|r^{12}| \approx 1$, whereas for a metal
bulk $|r^{12}|$ is small, so that waves generated by fluctuating source currents
in the bulk medium can propagate into and through the coating and therefore
contribute to the energy density in a much more significant way at distances
$z \gg d$. In fig.\ \ref{Fig:Dens_Bi_auf_GaN_bulk_coating} the numerical 
plots for Al/Bi and GaN/Bi systems confirm this expectation.  

\begin{figure}[Hhbt]
  \centering
  \begin{minipage}[t]{0.9\textwidth}
    \epsfig{file=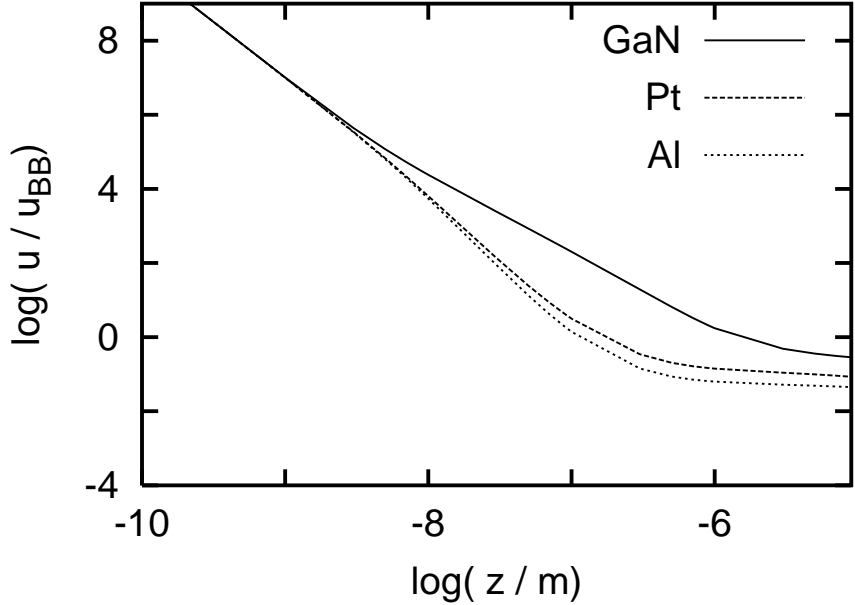, width = 0.9 \textwidth}
    \caption{Numerical results for the thermal near-field energy density $\langle u^{\rm total}_\parallel \rangle$
            above the layered structure for a substrate consisting of GaN, Pt or Al coated with a $5 {\rm nm}$ Bi
            layer at temperature $T = 300 {\rm K}$.}
    \label{Fig:Dens_Bi_tm}
  \end{minipage}
\end{figure}

Before finishing the discussion of the energy density, we give in
figs.\ \ref{Fig:Dens_Bi_te} and \ref{Fig:Dens_Bi_tm} two further numerically  
computed plots of $\langle u_\parallel^{\rm total} \rangle$ and $\langle u_\perp^{\rm total} \rangle$ for a $5 {\rm nm}$ Bi-coating
on different bulk materials. One sees that the $1/z^2$- and
$1/z^4$-power laws of $\langle u^{{\rm c}}_\parallel \rangle$ derived above leave their imprints  
in the TM-mode part of the total energy density. For the TE-mode part of the
energy density one has to distinguish between a metal-metal system
and a polar material-metal system, because for a metal-metal system
at $z \gg d$ the bulk contributions dominate the energy density, but for 
a polar bulk material this region is dominated by the contribution of
the metal coating only. 
 
\section{Surface plasmon coupling}
\label{Sec4}

The rise in the TM-mode part of the energy density for a polar substrate coated with a metal can 
be explained in terms of the low-frequency surface plasmon polariton resonance within
the coating. In the given geometry (see fig.\ \ref{Fig:Configuration}) the surface modes are given by the zeros of the
function~\cite{KliewerFuchs67,J.J.BurkeEtAl1986,H.Raether1980}
\begin{equation}
  N_\parallel = 1 - r_\parallel^{12} r_\parallel^{02} \re^{-2 \ri h_2 d} \equiv 0
  \label{Eq:dispersion_generell}
\end{equation}
with $h_2^2 = k_0^2 \epsilon_{r2} - \lambda^2$. This function coincides with the denominator of $T^{\rm c}_\parallel$ (cf.\  
$D_\parallel$ in eq.\ (\ref{Eq:Transmission_Coating})). For a non-magnetic material these surface modes are purely TM-polarized
and do exist for materials with a negative permittivity only~\cite{H.Raether1980}. For a polar substrate or bulk material
with a metal coating the Fresnel coefficient $r_\parallel^{12}$ can be approximated by $r_\parallel^{02}$ for all relevant 
frequencies. Within this rough approximation the dispersion relation in eq.\ (\ref{Eq:dispersion_generell}) coincides with the 
dispersion relation of a free standing metal film surrounded by a vacuum only. Therefore the conclusions drawn for a free
standing metal film~\cite{Biehs2007} can be applied to the metal-coated polar substrate. 

It follows~\cite{H.Raether1980,Biehs2007} that for coatings thinner than the skin depth $d_{\rm s}$ the two degenerate surface plasmon 
branches with the resonance frequency $\omega_s \approx \omega_{p}/\sqrt{2}$ split into two non-degenerate branches given by~\cite{H.Raether1980}
\begin{equation}
  \omega_\pm = \frac{\omega_{p2}}{\sqrt{2}} \sqrt{1 \pm \re^{-\lambda d}},
  \label{Eq:SPP_branches}
\end{equation}
where for convenience the plasma model is used to describe the permittivity. As expressed by eq.\ (\ref{Eq:SPP_branches}) the 
resonance frequency of the high-frequency surface plasmon polariton branch $\omega_+$ goes to the plasma frequency $\omega_{p2}$ 
of the coating, and the resonance frequency of the low-frequency branch $\omega_-$ goes to zero for very 
thin coatings, i.e., for $\lambda d \ll 1$. Due to the fact that the $\lambda$-integral for the energy density 
in eq.\ (\ref{Eq:Energy_density}) is dominated by lateral wave vectors of the order 
$\lambda \approx z^{-1}$, for $z \ll d$ the splitting of the surface plasmon branch cannot be observed, since
$\lambda d \gg 1$. In this case one obtains the same energy density as in the case of an infinitely thick coating. 
On the other hand, for observation distances $z \gg d$ in the near field above the coated material the surface 
plasmon coupling leads to a splitting of the surface plasmon branches, since
$\lambda d \ll 1$ in this case.
Therefore at these distances the resonance of the low-frequency branch $\omega_-$ will go to frequencies which 
are accessible thermally, leading to an increase in the thermal near-field energy density. 

\begin{figure}[Hhbt]
  \centering
  \begin{minipage}[t]{0.9\textwidth}
    \epsfig{file=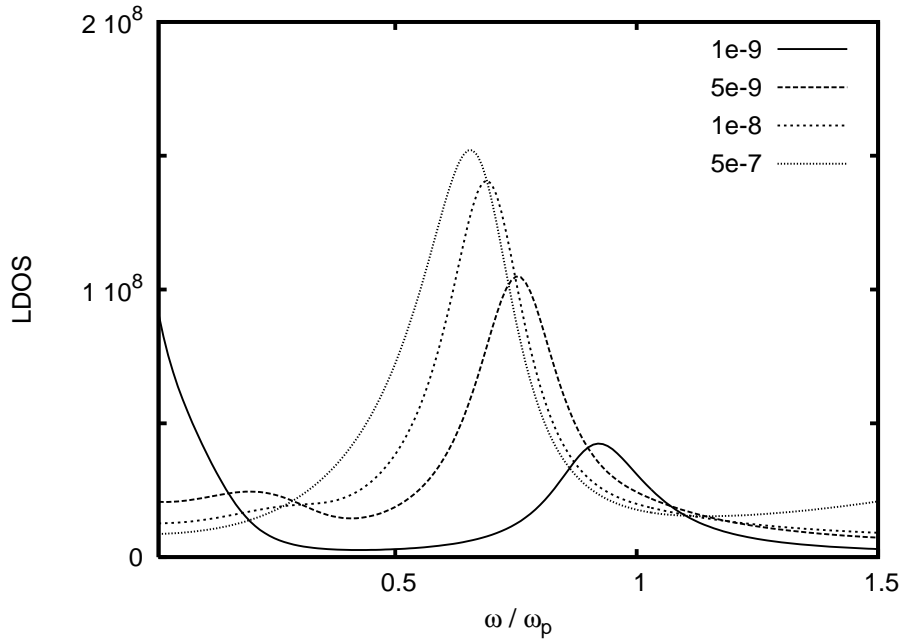, width = 0.9 \textwidth}
    \caption{Plot of the LDOS of the TM modes of a layered system for different thickness $d$ of a Bi-coating
             on a GaN-substrate. The frequencies are normalized to the plasma frequency of the coating material. }
    \label{Fig:LDOS_Bi_GaN}
  \end{minipage}
\end{figure}

In fig.\ \ref{Fig:LDOS_Bi_GaN} we plot the local density of states (LDOS) defined in~\cite{JoulainEtAl03} for the 
TM-modes only. One observes how the resonance at $\omega_s$ splits into two resonances, where the high-frequency resonance 
goes to $\omega_{p2}$ of the coating and the low-frequency resonance goes straight to zero. 
Thus, it reaches the thermally accessible region 
for thin coatings and increases the LDOS in that region, and therefore also the thermal near-field energy density 
leading to the $z^{-2}$-power law.

For a metal substrate coated with a metallic material, the dispersion relation in eq.\ (\ref{Eq:dispersion_generell}) can be 
approximated in the near-field region with $\lambda \gg k_0$ as
\begin{equation}
  \frac{\epsilon_{r2} - \epsilon_{r1}}{\epsilon_{r2} + \epsilon_{r1}} \frac{\epsilon_{r2} - 1}{\epsilon_{r2} + 1} \re^{-2 \lambda d} = 1
\end{equation}
which leads again within the plasma model to two surface plasmon polariton branches. In this case, for $z \gg d$, 
the resonance frequencies of the surface plasmon polariton branches go to the plasma frequency of the 
coating $\omega_{p2}$ and the surface
plasmon resonance frequency $\omega_{p1}/\sqrt{2}$ for arbitrarily thin coatings. Therefore the surface plasmon polariton
coupling will not lead to an increase of the LDOS in the thermally accessible region, since for real metals the plasma
frequencies are much greater than the thermal frequency $\omega_{\rm th} \approx 10^{14}{\rm s}^{-1}$ at $T = 300 {\rm K}$. 
It follows that the thermal near-field energy density is unaffected by the surface plasmon coupling, leading to values
below that of the semi-infinite body, and to a quite different $z^{-4}$-power law for metal substrates as previously shown
in fig.\ \ref{Fig:Dens_Bi_auf_GaN_bzw_Al_tm}. 

\begin{figure}[Hhbt]
  \centering
  \begin{minipage}[t]{0.9\textwidth}
    \epsfig{file=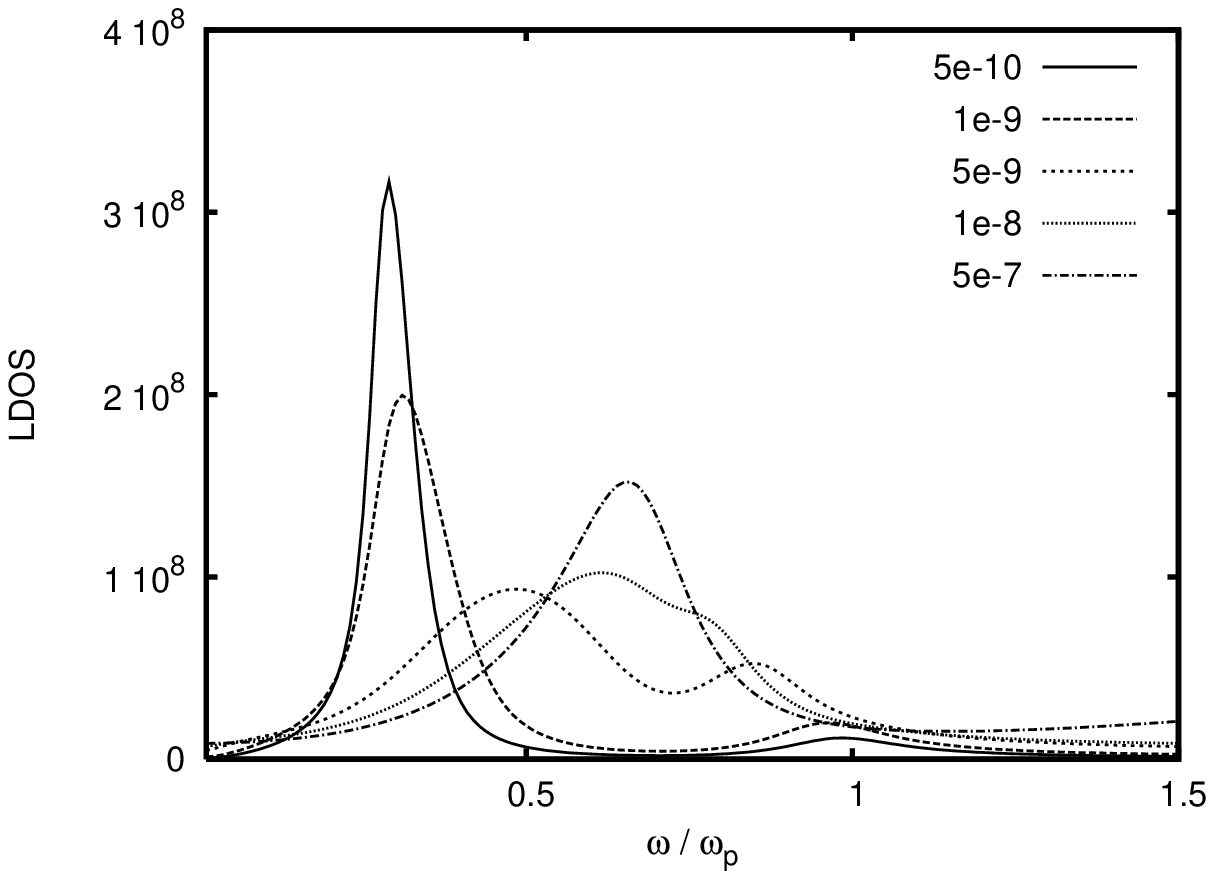, width = 0.9 \textwidth}
    \caption{Plot of the LDOS of the TM modes of a layered system for different thicknessess $d$ of a Bi-coating
             on a Pt-substrate. The frequencies are normalized to the plasma frequency of the coating material. }
    \label{Fig:LDOS_Bi_Pt}
  \end{minipage}
\end{figure}

In fig.\ \ref{Fig:LDOS_Bi_Pt} we plot the LDOS for the TM-modes for a Bi-coating on a Pt-substrate. 
The splitting of the surface plasmon resonance into two resonances is clearly visible. 
Here the high-frequency resonance goes to the plasma frequency of the coating and the low-frequency resonance
goes to the surface plasma resonance of the substrate given by 
$\omega_{p1}/\sqrt{2} = 8\cdot10^{15} {\rm s}^{-1} = 0.27 \omega_{p2}$ with 
$\omega_{p2} = 2.1\cdot10^{16}{\rm s}^{-1}$~\cite{AshcroftMermin76}.

\section{Thermal near-field radiation}
\label{Sec5}

\begin{figure}[Hhbt]
  \centering
  \begin{minipage}[t]{0.9\textwidth}
    \epsfig{file=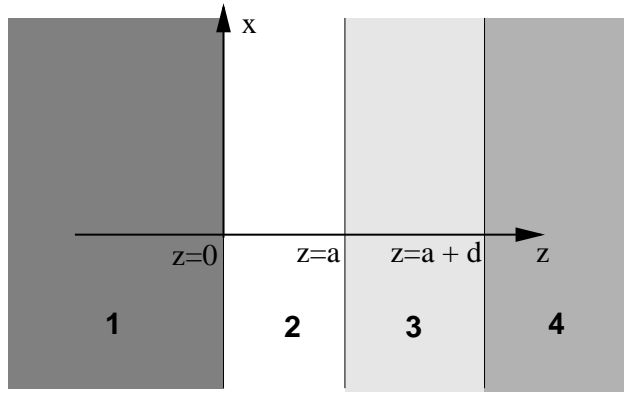, width = 0.6 \textwidth}
    \caption{Sketch of the configuration for the near-field radiative heat transfer between a semi-infinite medium at
             $z \leq 0$ and a coated semi-infinite medium at $z \geq a$.}
    \label{Fig:Configuration2}
  \end{minipage}
\end{figure}

In this last section we discuss the radiative near-field heat transfer between a semi-infinite body
and a coated semi-infinite body as sketched in fig.\ \ref{Fig:Configuration2}. Since the calculation follows the 
well-established rules, we proceed directly to the result for the Poynting vector in this geometry, assuming
$T_1 \neq 0$ for the material at $z < 0$ and $T_3 \neq 0$ for the layered structure at $z > a$. With $\epsilon_2 = \epsilon_0$,
the result takes the form 
\begin{equation}
\begin{split}
  \langle S_{z} \rangle &= \int\!\rd \omega\, \frac{E (\omega,T_1) - E (\omega,T_3)}{(2 \pi)^2} \biggl\{ \int_0^{k_0}\!\!\!\rd \lambda\, \lambda \frac{(1 - |r_\perp^{21}|^2)(1 - |R_\perp|^2)}{|N_\perp'|^2} \\
                                   &\quad + \int_{k_0}^\infty\!\!\!\rd \lambda\, \lambda \frac{4 \Im(r_\perp^{21}) \Im(R_\perp) \re^{- 2 \gamma a}}{|N_\perp'|^2} \, + \, \parallel \biggr\},
\end{split}
\label{Eq:Polder_van_Hove_Schicht_kap5}
\end{equation}
where the symbol $\parallel$ abbreviates the corresponding expressions for the TM-modes, and 
with the usual Fresnel coefficients $r_\perp$ and $r_\parallel$. In addition, 
\begin{equation}
  R = \frac{r^{23} + r^{34} \re^{2 \ri h_3 d}}{1 - r^{34} r^{32} \re^{2 \ri h_3 d}} \quad\text{and}\quad
  N' = 1 - r^{21} R \re^{2 \ri h_2 a}
\end{equation}
for TE- and TM-polarization, respectively. It can be easiliy checked that for $d \rightarrow \infty$ this expression
reduces to the Polder-van-Hove (PvH) result~\cite{PolderVanHove71} for the near-field radiative heat transfer 
between two semi-infinite bodies.

It is well-known~\cite{PolderVanHove71} that the radiative heat transfer between two metals described by the 
PvH expression~\cite{A.I.VolokitinB.N.J.Persson2001} is dominated by the TE-modes, whereas the radiative heat 
transfer between a metal and a polar material or two polar materials,
respectively, is dominated by the TM-modes giving 
\begin{equation}
  \langle S_\parallel \rangle \propto \frac{1}{a^2} \quad\text{and}\quad \langle S_\perp \rangle \propto {\rm const}
\end{equation}
in the near-field region. Hence, the exponents of the $1/z^3$- and $1/z$-dependence of the TM- and TE-mode 
parts of the thermal near-field energy density of a half-space are reduced by one. 
It is to be expected that the radiative heat transfer between a semi-infinite body and a layered structure with 
a thin coating of thickness $d \ll d_{\rm s}$ will again resemble the usual PvH expression for $a \ll d$, 
since the energy density above the layered structure coincides in this case with that of a semi-infinite body consisting 
of the coating material only. In the opposite case, for $a \gg d$, the radiative heat transfer should be 
determined by the change in the thermal near-field energy density described in
the preceding section. 

\begin{figure}[Hhbt]
  \centering
  \begin{minipage}[t]{0.9\textwidth}
    \centering
    \epsfig{file=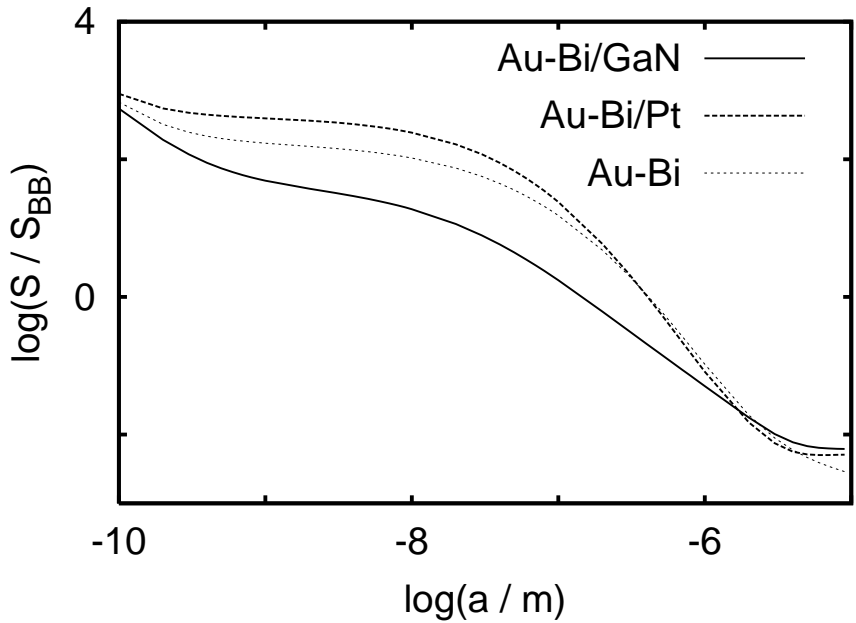, width=0.9\textwidth}
    \caption{Numerical results for the near-field radiative heat transfer between a semi-infinite Au-body
             with $T_1 = 300 {\rm K}$ and a coated semi-infinite GaN- or Pt-substrate with $T_3 = 0 {\rm K}$, 
             as functions of the gap width $a$.
             The thickness of the Bi-coating is chosen to be $5 {\rm nm}$.}
    \label{Fig:S_Halbraum_Au_Bi_GaN_bzw_Pt}
  \end{minipage}
\end{figure}

Furthermore one expects that when taking a metallic material for medium 1 the TE-modes of the layered 
structure dominate the heat transfer, so that the radiative heat transfer should behave similar to the thermal 
near-field energy density $\langle u_\perp^{\rm total} \rangle$ plotted in fig.\ \ref{Fig:Dens_Bi_te}. 
Choosing Au for medium 1 we get the near-field radiative heat transfer plotted in fig.\ 
\ref{Fig:S_Halbraum_Au_Bi_GaN_bzw_Pt}. Indeed this figure fully confirms this expectation.
Moreover, using a metal substrate such as Pt for medium 4, the
radiative heat transfer rises over the PvH-result for a Au-Bi configuration, as is explained by the contribution of
the Pt-substrate, so that in this case the radiative heat transfer in the layered structure 
is in prinicple, given by the PvH-result for a Au-Pt configuration. 

\begin{figure}[Hhbt]
  \centering
  \begin{minipage}[t]{0.9\textwidth}
    \centering
    \epsfig{file=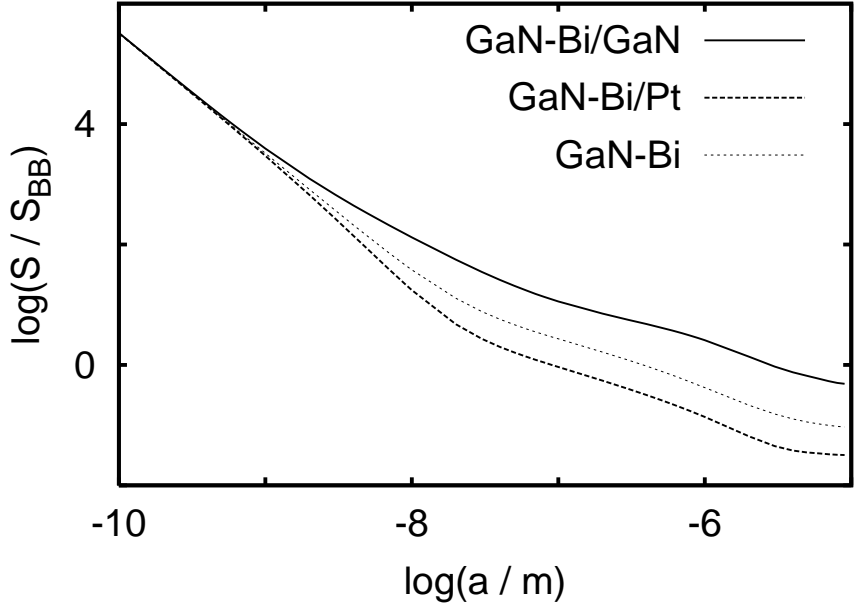, width=0.9\textwidth}
    \caption{Numerical results for the near-field radiative heat transfer between a semi-infinite GaN-body
             with $T_1 = 300 {\rm K}$ and a coated semi-infinite GaN- or Pt-substrate with $T_3 = 0 {\rm K}$,
             as functions of the gap width $a$.
             The thickness of the Bi-coating is chosen to be $5 {\rm nm}$.}
    \label{Fig:S_Halbraum_GaN_Bi_GaN_bzw_Pt}
  \end{minipage}
\end{figure}

On the other hand, choosing GaN for medium 1, we expect dominance of the TM-mode energy density depicted in 
fig.\ \ref{Fig:Dens_Bi_tm}, but with reduced power laws for $a \gg d$, i.e., the $1/z^2$-power law should 
lead to a radiative heat transfer proportional to $1/a$, whereas
the $1/z^4$-power law should lead to a radiative heat transfer proportional to $1/a^3$. This is exactly what is seen in the
numerical results plotted in fig.\ \ref{Fig:S_Halbraum_GaN_Bi_GaN_bzw_Pt}. Thus, it is possible to understand 
the near-field radiative heat transfer qualitatively from the thermal energy density of the considered materials. 
Even more interesting, the enhancement in the thermal near-field energy density due to surface plasmon polariton 
coupling in the coating material can be observed in the radiative heat transfer in a slab geometry as sketched in fig.\ \ref{Fig:Configuration2}.

\section{Conclusions}
\label{Sec6}

In this paper, we have given a discussion of the thermal radiation and the thermal near-field energy density 
of a metal-coated substrate. It has been shown that the maximum of the thermal radiation, which
is observed for free metal films at a certain thickness~\cite{Biehs2007}, does not appear for coated 
materials, since for thin coatings the thermal radiation of the substrate hides this maximum.

On the other hand, the increase in the thermal near-field energy density of a free standing metal film~\cite{Biehs2007}
due to surface plasmon polariton coupling inside the metal coating has also been found for a coated substrate, when a polar material 
is used as substrate. For metal coatings on metal substrates such an increase does not exist. Moreover, for metallic substrates the 
thermal near-field energy density $\langle u_\parallel^{\rm total} \rangle$ for observation distances $z \gg d$ and 
coating thickness $d \ll d_{\rm s}$ is some orders of magnitude smaller than for a polar substrate (with the same coating), 
obeying a rather different power law. This difference in behaviour resulting from the interchange
of the substrate material can be explained with the surface plasmon polariton coupling: For a polar substrate the thermally accessible
LDOS will be enhanced due to the low-frequency surface plasmon resonance, which goes to zero frequency for arbitrarily thin coatings, 
whereas for a metal substrate this resonance goes to the surface plasmon resonance of the substrate for arbitrarily thin coatings and
can therefore not be accessed thermally for plasma frequencies much greater than the thermal frequency.

In the last part we have shown that the differences investigated for the thermal near-field energy density of a coated material 
leave their imprints in the near-field radiative heat transfer between a semi-infinite body and a coated semi-infinite body.
Using a metal or a polar material allows one to 'select' the TE- or TM-mode part of the thermal
near-field energy density of the coated material to dominate the radiative near-field heat transfer. Therefore, it is possible to
observe the TM-mode-enhancement due to surface plasmon polariton coupling inside the coating by thermal heat transfer experiments.
Due to the fact that the expressions for the near-field radiative heat transfer and the vacuum 
friction~\cite{J.B.Pendry1997,A.I.VolokitinB.N.J.Persson2003} are fairly
similar, the discussed effect should also be observable for vacuum friction between coated materials.

Since a polarizable particle or an atom couples to the electric field, the radiative heat 
transfer~\cite{Dorofeyev98,MuletEtAl01,A.I.VolokitinB.N.J.Persson2002}, the spontaneous emission 
rate~\cite{J.M.Wylie1984,G.W.Ford1984,M.S.Tomas1995,H.T.DungEtAl2002} near a hot body and the thermal Casimir-Polder 
potential~\cite{P.Milonni1994,C.HenkelEtAl2002,A.I.VolokitinB.N.J.Persson2002} should be proportional 
to $\langle \mathbf{E}^2 \rangle \propto \langle u_\parallel \rangle$ in the near field, so that the discussed 
enhancement of the TM-mode part of the thermal near field should 
also enhance the near-field radiative heat transfer between a small particle and a coated material, the spontaneous emission rate
of an atom near a hot coated material, and the thermal Casimir-Polder potential, respectively. Moreover, the spin flip rate of 
atoms~\cite{C.HenkelEtAl1999,P.K.Rekdal2004} above a layered structure, which is in principle proportional 
to $\langle \mathbf{B}^2 \rangle \propto \langle u_\perp \rangle$, will also be changed by the use of thin coatings 
on appropriate substrates. Furthermore, it appears possible that the coherence of the thermal near 
field~\cite{CarminatiGreffet99,C.HenkelEtAl2000} can be controlled by the use of different metal coatings, since the surface
plasmon resonance frequency can be changed by the choice of the thickness of the coating. In this sense the 
discussion of the thermal energy density has a much broader field of application than the radiative heat transfer and the vacuum friction.

The author acknowledges support from the Studienstiftung des deutschen Volkes. Furthermore he thanks O. Huth, F. R\"uting, D. Reddig, and
M. Holthaus for helpful discussions and kind criticism.

\end{document}